\documentclass[%
 reprint,
 amsmath,amssymb,
 aps,
floatfix,
]{revtex4-2}

\usepackage{graphicx}
\usepackage{dcolumn}
\usepackage{bm}

\usepackage[paperwidth=210mm,paperheight=297mm,centering,hmargin=1.5cm,vmargin=2.0cm]{geometry}

\usepackage{siunitx}
\usepackage{empheq}
\usepackage{wrapfig}
\usepackage[usenames,dvipsnames]{xcolor}

\definecolor{cream}{RGB}{222,217,201}
\newcommand{\software}[1]{\textsc{#1}}
\newcommand{\flowrate}{\ensuremath{Q_{\mathrm{jet}}}}
\newcommand{\bnabla}{\ensuremath{\boldsymbol{\nabla}}}
\newcommand{\bcdot}{\ensuremath{\boldsymbol{\cdot}}}

\begin{document}

\preprint{APS/123-QED}

\title{Upstream motion of oil droplets in co-axial Ouzo flow due to Marangoni forces}

\author{Steffen~Bisswanger\textit{$^{ a}$}, Duarte~Rocha\textit{$^{ b}$}, Sebastian~Dehe\textit{$^{ a,c}$}, Christian~Diddens\textit{$^{ b}$}, Tobias~Baier\textit{$^{ a}$}, Detlef~Lohse\textit{$^{ b, d}$}, and Steffen~Hardt\textit{$^{ a, e, }$}\vspace{0.3cm}}

 \email{hardt@nmf.tu-darmstadt.de}

\affiliation{%
 \textit{$^{a}$~Technische Universität Darmstadt, Fachgebiet Nano- und Mikrofluidik, Peter-Grünberg-Str.~10, 64287 Darmstadt, Germany\\
 \textit{$^{b}$~Physics of Fluids Group and Max Planck Center for Complex Fluid Dynamics, Department of Science and Technology, J.M. Burgers Center for Fluid Dynamics, University of Twente, Enschede, The Netherlands}\\
 \textit{$^{c}$~ Linac Coherent Light Source SLAC National Accelerator Laboratory, Menlo Park, California 94025, United States}\\
 \textit{$^{d}$~ Max Planck Institute for Dynamics and Self-Organization, Am Faßberg 17, Göttingen, Germany}}
}%

\begin{abstract}
\vspace{0.6cm}
To explore the physicochemical hydrodynamics of phase-separating ternary liquids ("Ouzo-type"), a binary oil-ethanol mixture is introduced into a co-flowing stream of water. Oil droplets nucleate at the interface between the two liquids, leading to a larger oil droplet interacting with the ethanol-rich jet. Although buoyancy forces and hydrodynamic drag forces push the droplet in downstream direction, we observe an upstream motion. Using computational fluid dynamics simulations of a simplified model system, we identify the nucleation zone for oil droplets and uncover Marangoni forces to be responsible for the upstream motion of the droplet. A semi-analytical model allows us to identify the key parameters governing this effect. A general conclusion is that Marangoni stresses can reverse the motion of droplets through channels, where the surrounding liquid is a multi-component mixture. The insights from this work are not only relevant for channel flow, but more generally, for the physicochemical hydrodynamics of multiphase, multi-component systems.
\vspace{1.5cm}
\end{abstract}

\maketitle


\section{Introduction}
A vast number of technical applications involve flows of a liquid that consists of multiple components. Owing to differences in inter-molecular forces for different components, miscibility and interfacial tension heavily depend on the ratio of the different components. In single-phase systems, concentration gradients of dissolved components may influence the flow through changes in bulk properties (e.g., density), but they usually do not directly induce a flow. If, however, an interface is subject to a concentration gradient that results in an interfacial tension gradient, the tangential momentum balance at the interface tells us that it is inevitable that a flow is induced, commonly known as Marangoni flow~\cite{scriven1960marangoni, lohse2020physicochemical}. Depending on the cause of the interfacial tension gradient, this flow is categorized into thermal or solutal Marangoni flow. In turn, it can result in a deformation of the interface as is the case, e.g., for sessile droplets that spread or contract due to evaporation-induced solutal or thermal Marangoni stresses~\cite{baumgartner2022marangoni, karpitschka2017marangoni, kant2024autothermotaxis}, or in complex advection of droplets that are freely suspended in a stratified liquid~\cite{meijer2023rising, li2019bouncing, li2021marangoni, li2022marangoni}. As mentioned above, the composition of a liquid does not only affect interfacial tension, but also its miscibility. Probably the most famous and most studied example of such a system is a ternary mixture of water, ethanol and transanethole: For compositions with relatively large ethanol mass fraction and small water fraction, the three components are miscible. However, if we increase the water content, the three components will cease to remain fully miscible at some point, resulting in the nucleation of oil droplets. This spontaneous emulsification is often referred to as the "Ouzo effect"~\cite{vitale2003liquid, lohse2020physicochemical}, since the Greek spirit Ouzo (among many others) is one example of such a ternary system with a miscibility gap where emulsification can be observed. Typical droplet radii for "Ouzo type" emulsions are in the range of a few $\qty{100}{\,\nano\metre}$, depending on the component ratios, temperature and time~\cite{grillo2003small}. The resulting light scattering gives the emulsion its opaque appearance.  The dynamics of this system is especially interesting if we have large concentration gradients, as the creation of new interfacial area is particularly rapid. Additionally, large concentration gradients imply large interfacial tension gradients that can act on these interfaces. In combination, these two points imply the emergence of an emulsion in which the characteristic force (surface integral of the Marangoni stresses) per unit volume is relatively large, especially on small length scales~\citep{tan2019microdroplet}. This results in the occurrence of phenomena that are rich in dynamics, and to some extent counter-intuitive. In the context of applications in chemical analysis, the large ratio of interfacial area per unit volume in such "Ouzo-type" systems is leveraged in methods like dispersive liquid–liquid microextraction~\cite{rezaee2006determination, rezaee2010evolution, zgola2011dispersive}, where it enables a fast transfer of an analyte from the continuous aquous phase into the dispersed oil phase.\\\indent
In this paper, we study such an "Ouzo-type" system experimentally, numerically, and using analytical models: We create large concentration gradients by injecting an oil-ethanol mixture into a stream of pure water, following a similar approach as used by Hajian and Hardt~\cite{hajian2015formation} (see \S~\ref{sec:experiments}). Capturing the complex multi-scale dynamics occurring in this system through numerical simulations poses a substantial challenge. Resolving  the three-dimensional multiphase advection-diffusion problem, including growth and coalescence of droplets ranging from a few $\qty{100}{\nano\metre}$ to about $\qty{100}{\micro\metre}$ in diameter, is impractical. Instead, we are able to shine light onto specific aspects of the experimentally observed phenomena by making simplifying assumptions that preserve the system's fundamental physics (see \S~\ref{sec:simulations}). Using even further simplifying assumptions, we then rationalize the observations from the simulations with a semi-analytical model (see \S~\ref{sec:results}). The paper ends with conclusions and an outlook (see \S~\ref{sec:conclusion}).\vspace{-0.2cm}
\section{Experiments}\label{sec:experiments}
\begin{figure}[h!]
    \centering
    \includegraphics[width=8.6cm]{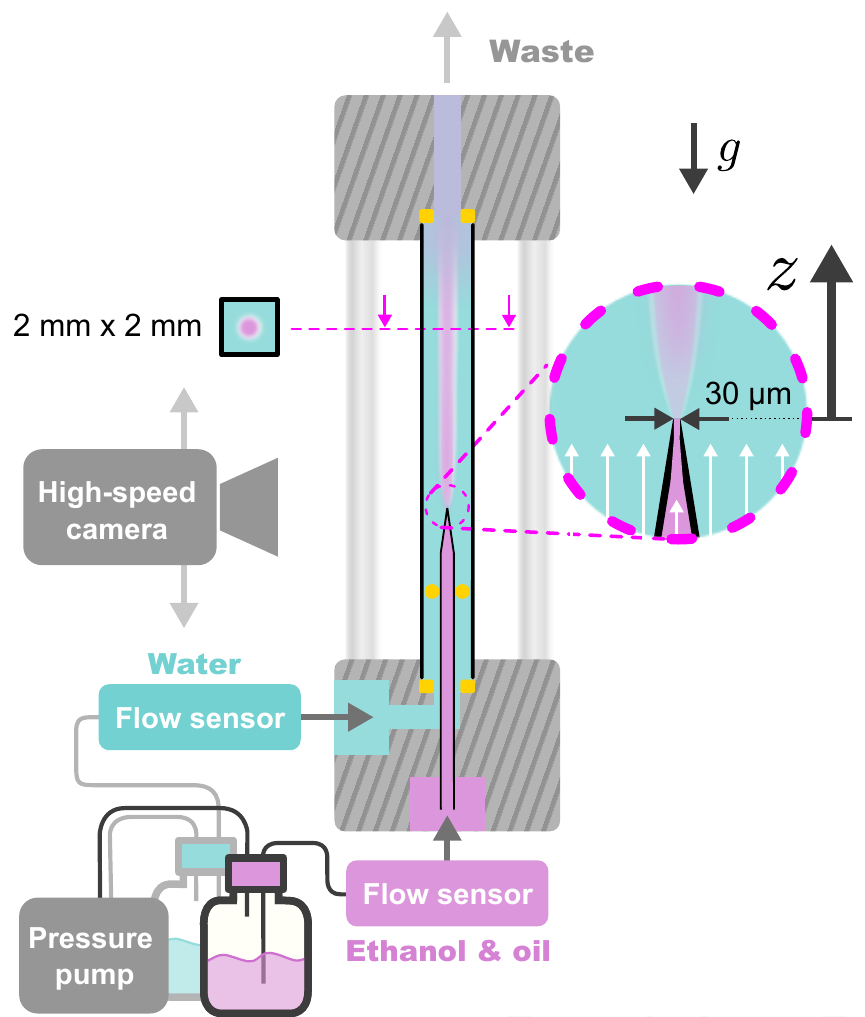}
    \caption{\label{fig:setup} Coaxial flow cell. The outer square capillary is clamped between two aluminum blocks using EPDM rubber (yellow) to create a seal. The inner circular capillary is centered using a circular EPDM rubber ring that is placed halfway on the inner capillary, allowing for the sheath flow to pass through at the edges of the outer capillary.}
\end{figure}

\begin{figure}[ht!]
    \centering
    \includegraphics[width=8.6cm]{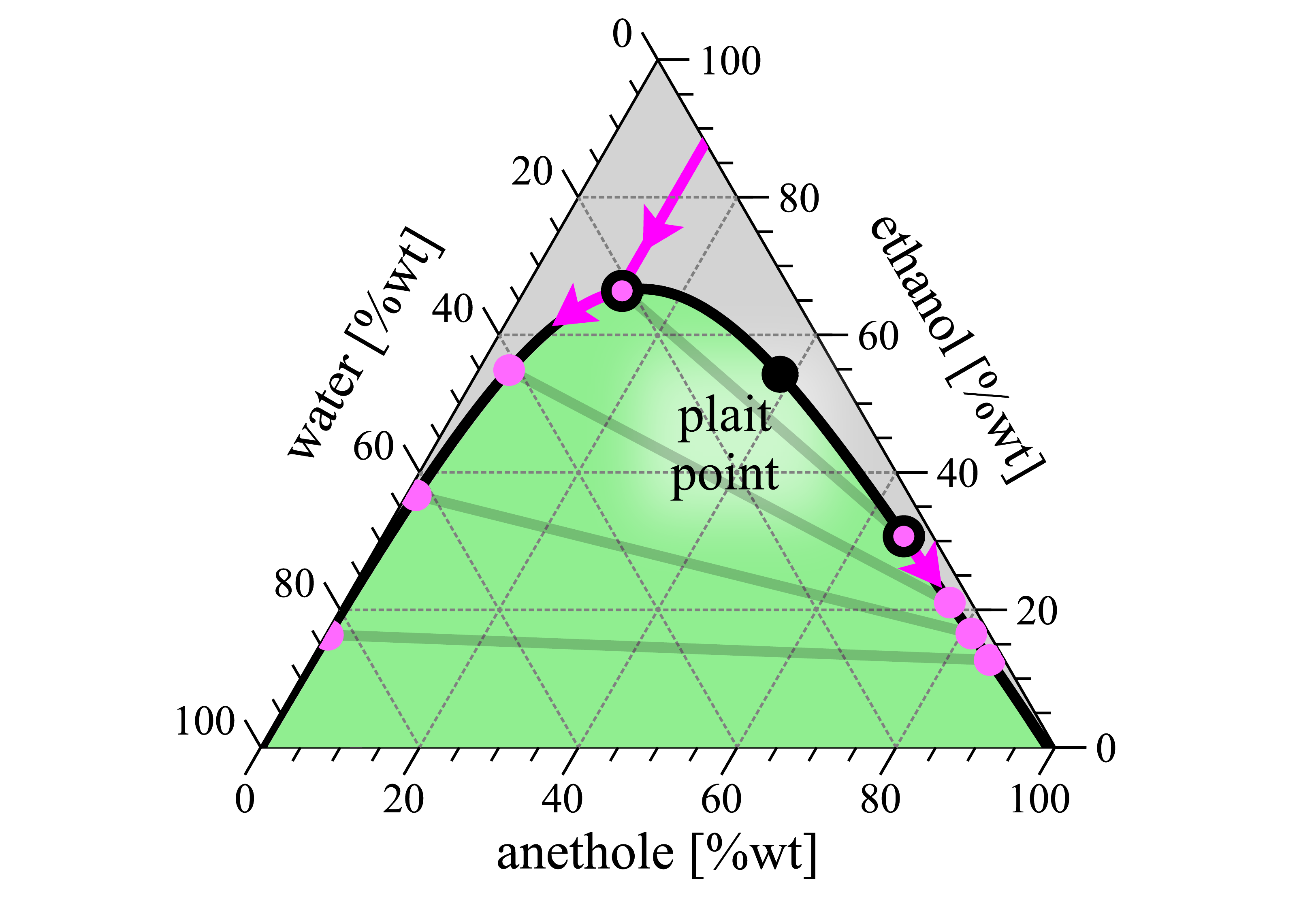}
    \caption{Ternary diagram of the Ouzo system in terms of mass fraction, as predicted by UNIFAC \citep{constantinescu2016further}. The solid black line denotes the binodal, and the dark green lines denote tie lines. The change in composition in the radial direction inside the flow is indicated with the magenta arrows, starting at a mixture of $\qty{12}{\percent}$ transanethole and $\qty{88}{\percent}$ ethanol by weight.}
    \label{fig:binodal_exp}
\end{figure}

\vspace{-0.2cm}
The experiment takes place in a vertical coaxial flow configuration (see Fig.~\ref{fig:setup}). A mixture of transanethole and ethanol (ratio of 12:88 by weight) is injected into a capillary with a sheath flow of water. Due to the radial gradients in concentration of the individual components where the sheath flow meets the jet, water diffuses inward, while ethanol and transanethole diffuse outward. While the binary mixture of the jet (ethanol-transanethole) is miscible for arbitrary ratios, the ternary mixture of water, ethanol, and transanethole is only fully miscible for compositions with small water fraction and large ethanol fraction. This means that there has to be an interval along the radial direction where we obtain a composition of the ternary mixture which at that point is not fully miscible anymore. This is qualitatively illustrated in the ternary diagram, as predicted by UNIFAC \citep{constantinescu2016further}, in Fig.\ \ref{fig:binodal_exp}: The composition along the center axis of the flow is given by the point near the top corner on the edge of the triangle, representing a binary mixture of ethanol and oil. As we move radially outward, the water content monotonically increases (ternary mixture). Very far outwards, the water fraction will eventually be unity (bottom left corner in Fig.\ \ref{fig:binodal_exp}). However, as the composition enters the two-phase region (green area in Fig. \ref{fig:binodal_exp}), the continuous phase follows the binodal curve along the water-rich side, while the excess oil separates into droplets. The corresponding composition of the oil-rich phase is then given by the point on the other side of the binodal connected by the corresponding tie-line (see Fig.~\ref{fig:binodal_exp}). The radial extent of this two-phase region surrounding the jet depends on the time available to diffusion to smear out the concentration field. Since we prescribe the concentration profile at some upstream position where we inject the binary mixture, the diffusion time can be mapped to the downstream position where the flow is observed. For this reason, further downstream the two-phase region is expected to become wider~\cite{hajian2015formation}.
\begin{figure*}
    \centering\includegraphics[width=17.8cm]{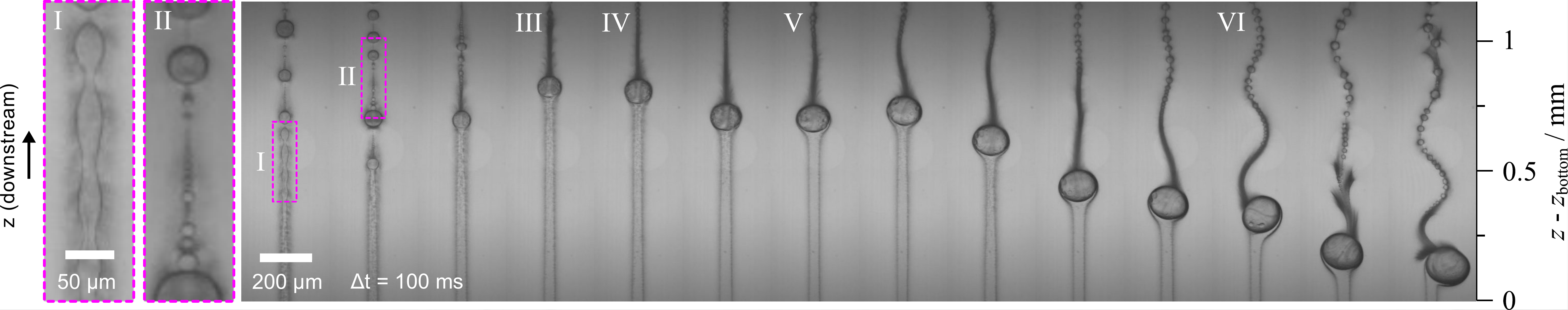}
    \caption{\label{fig:droplet_upstream} Image sequence showing the creation, growth and upstream motion of an oil droplet. On the left, a detailed view of the Rayleigh-Plateau-like instability (I) and the chain of droplets downstream of a large droplet (II) are shown. Once a droplet has formed, it grows while it moves downstream until it starts hovering (III), moves upstream (IV), gets deformed into an oblate shape and undergoes a rotational instability (V) that eventually transitions into a radial translational motion (VI). The maximum axial velocity of the droplet in upstream direction is $\qty{1.83}{\milli\metre\per\second}$. Sheath flow: \qty{175}{\micro\litre\per\minute} of pure water; Jet flow: \qty{5}{\micro\litre\per\minute} of 88:12 ethanol to transanethole, weight ratio. The bottom and the top of the images correspond to a downstream position of $z_{\mathrm{bottom}}=\qty{6.085}{\milli\metre}$ and $z_{\mathrm{top}}=\qty{7.238}{\milli\metre}$, respectively. The frames shown in this figure correspond to supplementary movie 1.}
\end{figure*}
\vspace{-0.2cm}
\subsection{Setup}\vspace{-0.2cm}
In order to implement the coaxial flow, we use a cell as depicted in Fig.~\ref{fig:setup}. The setup is oriented vertically so that buoyancy effects do not disturb the axisymmetry of the flow. It consists of an outer glass capillary with square cross-section (2 mm x 2 mm inner dimensions) and an inner glass capillary with circular cross-section that tapers down to a tip of 30 µm inner diameter. A closed-loop volume-flow-controlled pressure pump (Elveflow, model OB1 MkII equipped with two microthermal flow sensors) connected to the bottom of the cell creates a sheath flow of water (typically approx. $\qty{200}{\,\micro\litre\per\minute}$) and at the same time injects an ethanol-transanethole mixture through the inner capillary, thus creating a jet centered inside the sheath flow, rising as a buoyant plume. The ternary mixture exits the cell at the top ($z\approx\qty{4}{\centi\metre}$) and is directed into a waste container at atmospheric pressure. The flow inside the cell is imaged using a high-speed camera (Photron Fastcam Mini AX200) equipped with a long-distance microscope lens (Navitar 1-60191, 0.7-4x and Mitutoyo M Plan Apo 10x/0.28). Illumination (by means of an LED Panel) is provided from the backside. For cleaning the cell from transanethole droplets and from dust particles that can obstruct the flow inside the inner capillary nozzle, we flush the system with ethanol or immerse the dismantled system in 2\% Hellmanex III-water solution. Since the cleaning effect of the latter relies on surfactants that strongly affect interfacial tension, we thoroughly flush the cell with ethanol and water after using the Hellmanex III solution. The droplet growth in the freshly cleaned Ouzo system afterwards indicates the absence of surfactants \cite{nazarzadeh2013growth}.\vspace{-0.2cm}
\subsection{Creation and motion of nanodroplets}\vspace{-0.2cm}
In Fig.~\ref{fig:droplet_upstream}, a typical recording of the flow inside the experimental cell is shown. Along the entire length of the cell we observe a "dark mist" that surrounds the jet, which we interpret as nanodroplets nucleating in this region. As we go downstream, the thickness of this "misty region" does not vary much. The overall radial position, however, shifts inward at a typical speed of \qty{80}{\micro\metre\per\second} (\textit{cf.} supplementary information (SI)). At the same time, the jet velocity remains constant with respect to the axial position so that mass conservation implies that there is no radial net flow. Therefore, we interpret this as a relative motion between the nanodroplets and the surrounding continuous phase. Haijan and Hardt \cite{hajian2015formation} discussed this relative radial inward motion of nanodroplets in detail and identified Marangoni stresses as the driving force. As a result, the dispersed oil phase is focused close to the center axis.
\begin{figure*}
    \vspace{0.8cm}
    \centering\includegraphics[width=18.4cm]{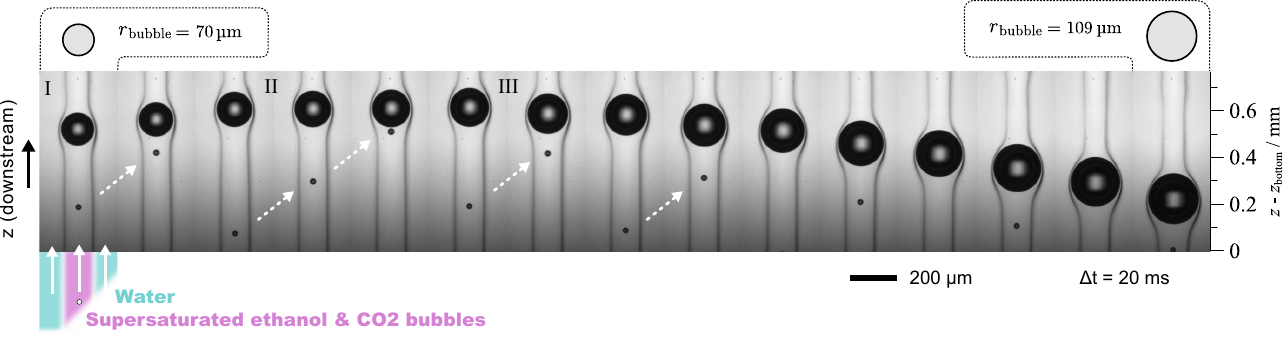}
    \caption{\label{fig:gas_bubble} Bubble size-dependent motion reversal of a \ensuremath{\mathrm{CO_2}} bubble in an ethanol jet in water. The ethanol jet is super-saturated with \ensuremath{\mathrm{CO_2}} and contains smaller bubbles (downstream motion indicated by dashed arrows) that coalesce with the large bubble. (I): bubble moves downstream. (II): bubble starts to hover at constant z-position. (III): bubble starts to move upstream. Throughout this process, the bubble grows due to coalesence with smaller bubbles and because of the flow of liquid supersaturated with \ensuremath{\mathrm{CO_2}} along the bubble surface. The flow rates are $\qty{200}{\micro\litre\per\minute}$ (water) and $\qty{23}{\micro\litre\per\minute}$ (ethanol). The bottom and the top of the images correspond to a downstream position of $z_{\mathrm{bottom}}=\qty{1.069}{\milli\metre}$ and $z_{\mathrm{top}}=\qty{1.834}{\milli\metre}$, respectively. The frames shown in this figure correspond to supplementary movie 2.}
    \vspace{0.2cm}
\end{figure*}
\vspace{-0.2cm}
\subsection{Formation of large droplets}\vspace{-0.2cm}
At positions sufficiently far downstream ($z>\qty{6}{\milli\metre}$, for the flow conditions studied here) the jet undergoes an instability that results in the periodic creation of droplets. While this system is without any doubt more complex than a single-component jet inside a second immiscible fluid, the droplet-shedding process is visually similar to the Rayleigh-Plateau instability, as shown in the detail view (I) of Fig.\ \ref{fig:droplet_upstream}. It is not easily observed when and how exactly the dense envelope of nanodroplets surrounding the jet coalesces and forms a continuous interface that separates the oil-rich inner phase and the water-rich outer phase, which could then be subject to the Rayleigh-Plateau instability. The diameter of the largest droplets formed in this process is approximately the same as the diameter of the jet. Between the larger droplets we observe smaller droplets that stay aligned with the center axis and subsequently coalesce into larger droplets (II).\vspace{-0.2cm}
\subsection{Upstream motion and instability}\vspace{-0.2cm}
Once a droplet is shed from the end of the jet, we often observe that its axial velocity is reduced to values below the centerline velocity of the surrounding flow, so that it moves relative to the jet in upstream direction. This effect quickly dies off if the jet upstream of the droplet is too unstable and fragments into more droplets. If, however, the droplet manages to move fast enough relative to the jet, as is the case in Fig.\ \ref{fig:droplet_upstream}, we can observe the droplet hovering at a fixed position (III), where it grows and eventually moves upstream (negative $z$-direction), while it keeps growing (IV). Since the end of the jet is then occupied by the droplet, no other large droplets can form behind the jet. Instead, a tail of small droplets can be observed downstream of the droplet. Further downstream, these droplets coalesce to form microdroplets that become individually visible. Once the oil droplet attached to the jet has become sufficiently large, it is notably deformed into an oblate shape (V). Up to this point, the configuration has remained fairly axisymmetric, at least with respect to the larger features of the flow. However, now the upstream-traveling droplet undergoes a second instability: First, we notice slight radial oscillations in the tail behind the droplet, indicating that the north pole of the droplet shifts in a rotational mode. These oscillations grow until eventually the droplet not only rotates but also translates in the radial direction (VI). As a result, the droplet describes an upstream meandering motion. The whole process can be observed in supplementary movie 1, where also the flow inside the droplet is qualitatively visible due to water-rich droplets that are trapped inside the oil droplet.  The upstream motion of the droplet continues until it reaches the tip of the inner capillary where it becomes pinned. Notably, we only observe a radial translational motion of the droplet (VI) in the focus plane, or at an angle $\qty{90}{\degree}$ relative to the focus plane. This means that the plane of translation is always aligned with the sidewalls of the outer capillary. A potentially similar aligning behavior is also known in the context of inertial focusing of particles inside microchannels with square cross-section\cite{zhou2013fundamentals}. However, the similarity in terms of the alignment does not allow for conclusions on the cause of the instability. The instability could still be caused by mechanisms unrelated to inertial focusing.\vspace{-0.2cm}
\subsection{Auxiliary experiment:\\Upstream motion of a gas bubble}\vspace{-0.2cm}
In order to better understand the conditions that are essential to the phenomenon of upstream motion of the oil droplet in the Ouzo system, we conduct a simplified variant of the experiment. We replace the oil-ethanol mixture in the Ouzo experiment with pure ethanol that is saturated with \ensuremath{\mathrm{CO_2}} at an over pressure of approximately 1 bar as the jet flow. By adding a large flow resistance into the inner capillary a bit upstream of where the ethanol is injected into the sheath flow of water, we create a large pressure drop that allows the dissolved \ensuremath{\mathrm{CO_2}} to start to come out of solution only as it exits the nozzle. The outer flow is pure water as in the original Ouzo experiment. Depending on the nucleation sites at the inner capillary wall, bubbles of various sizes can be produced that are flushed with the ethanol into the sheath flow of water. In Fig. 4 time series images of this experiment is shown where a \ensuremath{\mathrm{CO_2}} bubble rises inside the ethanol jet that is comparable in size to the oil droplet in the Ouzo system. Similar as the oil droplet in the Ouzo system, the gas bubble grows due to coalescence with small gas bubbles from the upstream side and due to the mass flow of \ensuremath{\mathrm{CO_2}} through the surface of the bubble. As for the oil droplet, this coincides with a downstream motion of the bubble (I), followed by hovering (II) and an upstream motion (III), a sequence of processes during which the bubble increases in size. This suggests that the explanation for the upstream motion of the droplet in the Ouzo system can be found by resorting to simplified two-phase models. In a more general context, this auxiliary experiment demonstrates that the motion reversal is also possible in a system where not only drag forces but also buoyancy forces have to be overcome.\vspace{-0.2cm}

\section{Numerical Simulations}\label{sec:simulations}\vspace{-0.2cm}
The numerical simulations were designed to show the fundamental mechanisms behind some key findings from the experiments. 
Given the system's inherent complexity -- which includes nanodroplet nucleation, Marangoni stresses at the complex boundaries between different phases, as well as mass transfer within the ternary mixture -- fully transient, full-scale simulations are impractical. 
Instead, we isolate distinct phenomena by employing simplified models that facilitate a focused investigation.\vspace{-0.2cm}

\subsection{Methods}\vspace{-0.2cm}

The simulations are performed with the open-source sharp-interface arbitrary Lagrangian–Eulerian (moving mesh) finite element software package \software{pyoomph} \cite{diddens2024bifurcation}, which builds on \software{oomph-lib} \cite{heil2006oomph} and \software{GiNaC} \cite{bauer2002introduction}.
All domains are meshed using unstructured triangular Lagrangian elements.

We model a cylindrical domain with a cross-sectional area equal to that of the outer square capillary used in the experiments. This approximation is justified, as the outer wall is sufficiently distant from the region of interest such that the flow dynamics is insensitive to the precise cross-sectional shape. A central tapered capillary, with \SI{30}{\micro\meter} diameter, injects a jet of fluid -- at a volumetric flow rate \flowrate -- into a surrounding sheath of pure water, maintained at a constant flow rate of ${Q}_{\mathrm{water}} = \SI{175}{\micro\liter\per\minute}$, as in the experiments. 
All calculations are performed on a 2D mesh using a cylindrical coordinate system, assuming axisymmetry.

The fluid motion is governed by mass conservation and the Navier-Stokes equations
\vspace{0.1cm}
\begin{equation}
\partial_t \rho + \bnabla \bcdot (\rho \mathbf{u}) = 0,
\end{equation}
\begin{equation}
\rho \left( \frac{\partial \mathbf{u}}{\partial t} + \mathbf{u} \bcdot \bnabla \mathbf{u} \right) = -\bnabla p + \bnabla \cdot \left(\mu (\bnabla \mathbf{u} + \bnabla \mathbf{u}^T)\right) + \rho \mathbf{g},\vspace{0.2cm}
\end{equation}
where $\mathbf{u}$ denotes the velocity, $p$ is the pressure, $\rho$ the density, $\mu$ the dynamic viscosity, and $\mathbf{g}$ the gravitational acceleration. 
The material properties depend on the local composition, which is in turn described by the mass fractions $w_{\mathrm{EtOH}}$ and $w_{\mathrm{H_2O}}$, while that of transanethole is given by
$w_{\mathrm{oil}} = 1 - w_{\mathrm{EtOH}} - w_{\mathrm{H_2O}}$.
The transport of each active component $i$ (either ethanol or water) is modeled by an advection-diffusion equation:
\vspace{0.1cm}
\begin{equation}
\rho \left( \frac{\partial w_i}{\partial t} + \mathbf{u} \bcdot \bnabla w_i \right) = \bnabla \cdot \left( \rho D_i \bnabla w_i \right),\vspace{0.2cm}
\end{equation}
with $D_i$ being the diffusion coefficient for component $i$. While the full diffusion matrix is generally non-diagonal, cross-diffusion effects are neglected in this study for simplicity. 
The liquid properties of the ternary mixture are approximated using a weighted average of the properties of transanethole oil at \SI{25}{\celsius} and experimentally fitted data for water–ethanol mixtures. 
The fitted mass density and dynamic viscosity values are taken from Ref.~\citep{gonzalez2007density}, and the diffusion coefficient is sourced from Ref.~\citep{par2013mutual}. 
The nucleation of oil droplets is governed by the thermodynamic stability of the ternary mixture. The UNIFAC model \citep{constantinescu2016further} is used here for phase equilibria modeling.

In \S~\ref{sec:nucleation_zones}, we first perform a single-phase simulation under the experimental conditions to identify regions where the Ouzo effect triggers nanodroplet nucleation. 
For these simulations, the inner jet has a flow rate of $\flowrate=$~\SI{5}{\micro\liter\per\minute} and an ethanol mass fraction of $w_{\mathrm{EtOH}}= $~\SI{88}{\percent}.
The quasi-stationary solution of the composition allows us to identify the regions where the Ouzo effect is expected to occur, based on the binodal curve (\textit{cf.} Fig.\ \ref{fig:binodal_exp}).

In \S~\ref{sec:axial_motion}, we consider a simplified two-phase model where a pure transanethole droplet is suspended in a binary ethanol-water mixture. 
Focusing solely on hydrodynamic interactions and neglecting mass transfer in between the two phases, we study the regime in which the net force on a droplet -- arising from Marangoni stresses, buoyancy, and advection -- vanishes, keeping the droplet fixed at $z_{\mathrm{drop}}=$\SI{6.5}{\milli\meter}. By assuming no mass transfer, we consider that once the system is close to hydrodynamic equilibrium, it is also near chemical equilibrium, and the limited mass exchange between phases has a negligible effect on the overall flow behavior.
By varying the droplet radius $R_{\mathrm{drop}}$ and the ethanol jet flow rate $\flowrate$, we determine the conditions that drive the droplet upstream or downstream.
We compute the critical flow rate $\flowrate^{\mathrm{crit}}$, defined via a Lagrange multiplier enforcing zero net force, that leads to a quasi-stationary equilibrium of the system.
At the interface between a droplet and the continuous phase (host fluid), the dynamic boundary condition reads
\vspace{-0.05cm}
\begin{equation}
\mathbf{\tau} ^h \bcdot \mathbf{n} - \mathbf{\tau} ^d \bcdot \mathbf{n} = \gamma \mathcal{K} \mathbf{n} + \bnabla_t \gamma,\vspace{0.2cm}
\end{equation}
where $\mathbf{\tau}^\phi = - p^\phi \mathbf{I} + \mu^\phi \mathbf{n} \bcdot \bigl( \bnabla \mathbf{u} + (\bnabla \mathbf{u})^T \bigr)$ represents the stress tensor of the phase $\phi$ (either host fluid, $h$, or droplet, $d$) at the droplet interface, $\gamma$ is the interfacial tension, $\mathcal{K}$ is the curvature of the interface, and $\mathbf{n}$ and $\mathbf{t}$ are the unit normal and tangential vectors, respectively. The interface is allowed to deform depending on the local interfacial tension. The interfacial tension is made composition-dependent according to a fit to the experimental data (see figure 5 in Ref. \cite{archer2024experimental}). We assume that although a high ethanol mass fraction could in principle dissolve the transanethole droplet, the droplet remains intact and its interfacial tension is set by the fit to the experimental data of \citet{archer2024experimental}.

Overall, this model framework enables us to capture the essential physics of the axial motion of oil droplets in the jet under a range of conditions.
Again, naturally, the simplified models do not account for all the complexities of the experimental system, which prevents quantitative comparisons.
\vspace{-0.2cm}
\subsection{Nucleation zones around the Ouzo jet}\label{sec:nucleation_zones}
\begin{figure}[ht!]
    \centering
    \includegraphics[width=8.6cm]{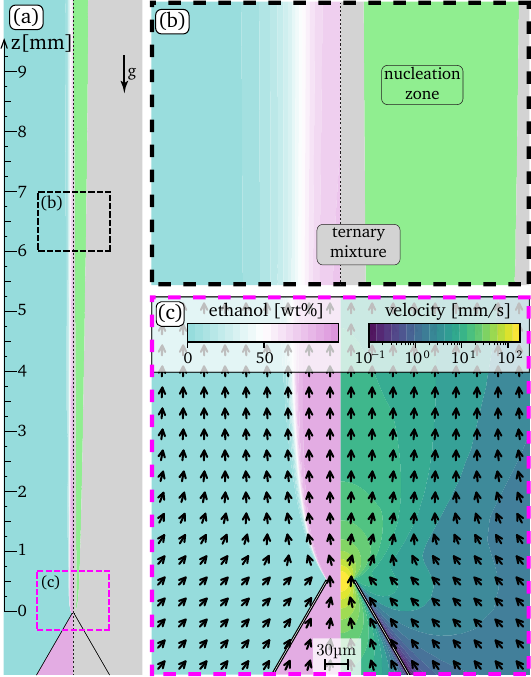}
    \caption{
    Quasi-stationary solution for the single-phase system under the experimental conditions.
    (a) Composition and jet development, with gravity acting from top to bottom. 
    At the left, the ethanol mass fraction ($wt\%$) is shown, 
    at the right the regions prone to nucleation (green) based on the binodal curve from Fig.\ \ref{fig:binodal_exp} are indicated.
    At the nozzle, an ethanol-transanethole mixture enters through the inflow boundary, while water flows in at the outer boundary.
    (b) Regions prone to nucleation due to Ouzo effect (right) and ethanol $wt\%$ (left) within an area between \SI{6}{\milli\meter} and \SI{7}{\milli\meter} away from the nozzle.
    (c) Ethanol $wt\%$ (left) and velocity magnitude (right) close to the nozzle.
    }
    \label{fig:jet_width}
    \vspace{-0.5cm}
\end{figure}

\vspace{-0.2cm}
Under the experimental conditions, the injected ethanol-transanethole jet gradually diffuses into the surrounding water as it travels downstream, creating peripheral regions where nucleation is favored according to the binodal curve (see Fig.~\ref{fig:binodal_exp}). 
Fig.\ \ref{fig:jet_width}a illustrates the evolving ethanol concentration in the whole domain (left), while highlighting the zones prone to nucleation in green (right). 
In Fig.~\ref{fig:jet_width}b, we show the areas susceptible to the Ouzo effect at a distance from \SI{6}{\milli\meter} to \SI{7}{\milli\meter} from the nozzle.
Fig.~\ref{fig:jet_width}c shows further details on the ethanol mass fraction (left) and the velocity magnitude (right) in the region close to the nozzle.\vspace{-0.2cm}

\vspace{-0.2cm}
\subsection{Axial motion of a droplet in the ethanol jet}\label{sec:axial_motion}\vspace{-0.2cm}

\begin{figure}[h]
    \centering
    \vspace{1cm}
    \includegraphics[width=8.8cm]{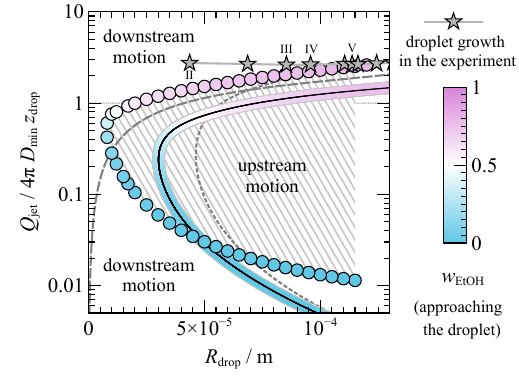}
    \caption{
        Critical jet flow rate $\flowrate^{\mathrm{crit}}/{4\pi D_\text{min} z_{\mathrm{drop}}}$ as function of droplet radius $R_\mathrm{drop}$ where the droplet remains stationary at a fixed position with ${z_{\mathrm{drop}}=\qty{6.5}{\milli\metre}}$ and $D_{\mathrm{min}}=\qty{0.365e-9}{\square\metre\per\second}$. The circular markers represent the numerical results (see \S~\ref{sec:simulations}). The solid black line corresponds to the estimate from the further simplified model described in \S~\ref{sec:results}, where the color gradient around the line indicates the mass fraction of ethanol approaching the droplet. The dashed lines correspond to the limiting cases described by equation (\ref{eq:largeQ}) (coarse dashed) and equation (\ref{eq:smallQ}) (fine dashed, see \S~\ref{sec:results}). The colorbar indicates the ethanol mass fraction of the liquid approaching the droplet (taken at $r=0$ and $z=z_{\mathrm{drop}}-2R_{\mathrm{drop}}$). The hatched region indicates the region where the droplet is expected to move upstream, according to the numerical simulation results. In the solid white region it moves downstream. The star symbols correspond to the flow rate and droplet growth in the experiment shown in Fig. \ref{fig:droplet_upstream}, with corresponding Roman numbering, for general orientation only, without the intent to imply a quantitative comparison or to indicate the axial motion in the experiment.}\label{fig:comparison}
        
\end{figure}

Using the simplified two-phase model, we study the axial motion of the droplet within the ethanol jet. 
By varying the droplet radius, $R_\mathrm{drop}$, from \SI{15}{\micro\meter} to \SI{120}{\micro\meter}, we determine the critical flow rate, $\flowrate^{\mathrm{crit}}$, at which the droplet remains at a constant downstream position $z_{\mathrm{drop}} = \qty{6.5}{\milli\metre}$. 
Fig.~\ref{fig:comparison} shows the numerical results (markers) alongside a theoretical estimate (solid line) which will be derived in \S~\ref{sec:results} from a force balance between Marangoni forces and viscous drag.
In our analysis, the critical flow rate is non-dimensionalized by $4\pi D_\text{min}\,z_{\mathrm{drop}}$, where $D_\text{min}$ is the minimum diffusivity of ethanol in water and $z_{\mathrm{drop}}$ is the fixed axial position of the droplet. 
Detailed justification for this scaling is provided in \S~\ref{sec:results}, where we also discuss the stability of the equilibrium position with respect to axial perturbations.
In the SI's \S S2, the flow features of four selected sets of parameters within the phase space of Fig.~\ref{fig:comparison} are shown.

Notably, the results reveal two distinct branches corresponding to small and large flow rates. 
The presence of multiple equilibrium flow rates for a given droplet radius is linked to variations in the ethanol concentration, as indicated by the color gradient of the markers in Fig.~\ref{fig:comparison}.
\vspace{-0.2cm}
\section{Discussion and a simple theoretical model}\label{sec:results}\vspace{-0.2cm}
Qualitatively, the nucleation zone around the jet (see \S~\ref{sec:nucleation_zones}) looks similar to the "dark mist" around the jet from the experiments, indicating that what we see are in fact nanodroplets. 
One aspect that the numerical single-phase simulations do not capture by design is the fact that the nanodroplets that nucleate migrate radially inward due to Marangoni stresses.

The simulations on the axial motion of the droplet (see \S~\ref{sec:axial_motion}) show that droplets can hover at a fixed axial position, as observed in the experiments. In this context, we can identify the experimental regime as belonging to the top branch shown in Fig. \ref{fig:comparison}, as indicated with the star symbols. We do not observe the bottom branch in experiments. This is due to two reasons: first, the equilibrium of the droplet in the bottom branch is unstable with respect to axial perturbations, as we will soon see. Second, the observation of a hovering droplet requires the creation of a sufficiently large droplet in the first place. This however only happens for large flow rates. Irrespective of that, we can rationalize why we observe these two branches in the simulations.\\ \noindent
\subsection{\hspace{-0.2cm}Equilibrium configuration of the hovering droplet}
The fact that we observe two different equilibrium configurations for a single droplet size indicates that the explanation for this has to be found in the approaching jet that is independent of the hovering droplet. According to Crank \cite{crank1979diffusion}, the mass fraction of ethanol at the centerline of the jet is approximately given by
\begin{equation}
    w_{\mathrm{EtOH}}(t)|_{{r=0}} = 1 - \exp{\left(\frac{-r_{0}^{2}}{4Dt}\right)},
    \label{eq:wcenter1}
\end{equation}
where $r_{0}$ is the radius that ethanol is initially confined to, $D$ is the diffusivity that is assumed to be constant here, and $t$ is the time elapsed along the center-streamline from the nozzle to the downstream position $z$. In a frame-of-reference moving at the centerline flow velocity $u_{\mathrm{c}}$, and by using $Q_{\mathrm{jet}} = \pi{}r_{0}^{2}u_{\mathrm{c}}$, we can rewrite this as
\begin{equation}
    w_{\mathrm{EtOH}}(z)|_{{r=0}} = 1 - \exp{\left(\frac{-Q_{\mathrm{jet}}}{4\pi{}Dz}\right)}.
    \label{eq:wcenter2}
\end{equation}
\begin{figure}[ht!]
    \centering
    \includegraphics[width=8.6cm]{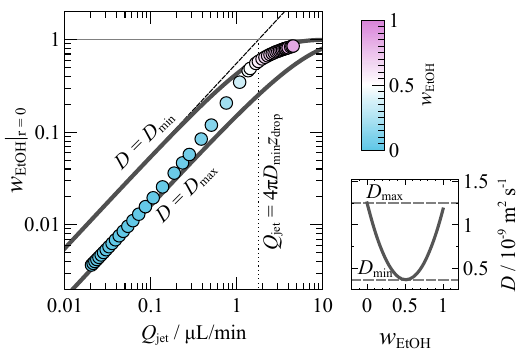}
    \caption{
        Mass fraction of ethanol at the centerline $w_{\mathrm{EtOH}}|_{r=0}$ as a function of the jet flow rate $Q_{\mathrm{jet}}$ (left) and diffusion coefficient of a binary water-ethanol mixture as function of the ethanol mass fraction (bottom right). The markers (left) represent the numerical results, with the colormap being identical to that of Fig. \ref{fig:comparison}. The solid black lines (left) correspond to equation~(\ref{eq:wcenter2}), with ${z=z_{\mathrm{drop}}}$, and the cases of maximum and minimum diffusion coefficient of water-ethanol mixtures ${D_{\mathrm{min}}=\qty{0.365e-9}{\square\metre\per\second}}$ and ${D_{\mathrm{max}}=\qty{1.255e-9}{\square\metre\per\second}}$. The dashed line (left) represents ${Q_{\mathrm{jet}} \cdot 4\pi{}D_{\mathrm{min}}z_{\mathrm{drop}}}$.
        }
        \label{fig:centerline_concentration}
\end{figure}
The agreement of this expression with the simulations is shown in Fig.\ \ref{fig:centerline_concentration}, revealing that $w_{\mathrm{EtOH}}$ at $z=z_{\mathrm{drop}}$ is well-described for small $Q_{\mathrm{jet}}$ using the maximum diffusion coefficient $D_{\mathrm{max}}$, while for large $Q_{\mathrm{jet}}$ it is well-described using the minimum diffusion coefficient $D_{\mathrm{min}}$. Generally, the diffusion coefficient of ethanol in water is concentration-dependent. In our case, the concentration does vary across the entire possible range, since the local ethanol mass fraction in our problem can vary between zero (outer flow of pure water) and unity (inner flow of pure ethanol). This, in turn, affects how the centerline concentration at a specific axial position $z$ varies with the flow rate, which is evident in the comparison. One aspect to note here is that even though the diffusion coefficient is only minimal for an ethanol mass fraction of approx. $0.5$ (see small plot in Fig.~\ref{fig:centerline_concentration}), the minimum diffusion coefficient also determines the centerline concentration for large flow rates, where the centerline concentration itself is close to unity. This is due to the fact that the radial diffusion problem is dominated by the minimum diffusivity along the radial direction, which for centerline concentrations above 0.5 is the global minimum of the diffusivity. We use this minimum diffusivity to non-dimensionalize the flow rate, as shown in Fig.\ \ref{fig:comparison}. The magnitude of the non-dimensional flow rate in relation to unity is what we refer to as large or small jet flow rates.\\\indent
In order to further explain the two different branches of equilibrium we now employ a force balance between the Marangoni force and the viscous force. Due to the complexity of the system, we do not aim at quantitative predictions, but instead at preserving the fundamental physics while making strongly simplifying assumptions. Based on a conservative estimate ($\qty{1.46}{\milli\metre\per\second}\geq u_{\mathrm{c}}\geq \qty{8}{\milli\metre\per\second}$, $\qty{8}{\micro\metre} \geq R_{\mathrm{drop}} \geq \qty{120}{\micro\metre}$, $\qty{1}{\milli\pascal\second} \geq \mu \geq \qty{2}{\milli\pascal\second}$ and $\qty{800}{\kilo\gram\per\cubic\metre} \geq \rho \geq \qty{998}{\kilo\gram\per\cubic\metre}$), the range of the Reynolds number is $0.009 < Re < 2$.\\
We approximate the viscous force as Stokes drag~\cite{stokes1851effect}
\begin{equation}
    F_{\mathrm{drag}} = 6\pi\mu\beta{}u_{\mathrm{c}}R_{\mathrm{drop}},
    \label{eq:Fdrag}
\end{equation}
where $R_{\mathrm{drop}}$ is the droplet radius and $\beta$ is a dimensionless factor that describes the increased flow velocity due to Marangoni flow. The centerline velocity $u_{\mathrm{c}}$ only gives a lower bound for the characteristic relative velocity between the droplet and the surrounding liquid flow (relevant for the drag force). In our case, the relative velocity increases due to Marangoni convection, so that in general $\beta\geq1$. The value of $\beta$ is \textit{a priori} unknown for a specific flow condition and might vary with specific parameters. However, as we will see, for the limiting cases, both $\beta$ and $u_{\mathrm{c}}$ will drop out due to their additional effect on the diffusion problem around the droplet.  The prefactor $6\pi$ in equation~(\ref{eq:Fdrag}) corresponds to the drag exerted on a sphere with a no-slip boundary condition at low Reynolds numbers (Stokes drag). In the Hadamard–Rybczynski solution, which describes the drag exerted on a droplet for small Reynolds numbers, we have a prefactor of $5\pi$ (assuming that the viscosity ratio between the droplet and the surrounding phase is unity)~\cite{hadamard1911mouvement,rybczynski1911uber}. Neither of these two cases describes our flow problem exactly due the qualitative effect that the Marangoni stresses have on the flow field. However, the following analysis is applicable irrespective of the exact choice of the drag coefficient in equation~(\ref{eq:Fdrag}), which does not affect the qualitative validity of the result.\\\indent
A comparison of the magnitude of the drag force with the magnitude of the buoyancy force acting on the droplet shows that we can neglect the effect of buoyancy: The ratio of these two forces can be approximated as
\begin{equation}
\frac{F_{\mathrm{buoyancy}}}{F_{\mathrm{drag}}} = \frac{4/3\pi{}R_{\mathrm{drop}}^{3}\Delta\rho}{6\pi\mu\beta{}u_{\mathrm{c}}R_{\mathrm{drop}}},\label{eq:buoyancy}
\end{equation}
where $\Delta\rho$ is the mass density difference between the oil droplet and the surrounding liquid. We can estimate the upper bound of this ratio by considering the interval of droplet radii covered in the numerical simulations in \S~\ref{sec:simulations}, where we have ${R_{\mathrm{drop}}\leq\qty{120}{\micro\metre}}$,  ${\left|\Delta\rho\right| \leq \qty{0.2}{\kilo\gram\per\cubic\metre}}$, ${\mu\geq\qty{1e-3}{\pascal\second}}$, ${\beta\geq1}$ and ${u_{\mathrm{c}}\geq\qty{1.4e-3}{\metre\per\second}}$. This gives 
\begin{equation}
\left|\frac{F_{\mathrm{buoyancy}}}{F_{\mathrm{drag}}}\right| < \qty{5e-4}{}
\end{equation}
and we therefore neglect the buoyancy force acting on the droplet in the further analysis. Note that we do not neglect the effect that buoyancy has on the rise velocity of the ethanol jet inside water at this point. This dependence enters our analysis through $u_{\mathrm{c}}$. Due to the fact that the ethanol jet is buoyant, $u_{\mathrm{c}}$ can be larger than its estimate based on a purely pressure-driven flow (see supplementary information \S~S3.3 for details).\\\indent
In order to estimate the Marangoni force and in order to simplify the diffusion problem, we approximate the droplet as a cylinder with the same radius and height as the droplet radius, as shown in Fig.~\ref{fig:kinematics}. This constrains the cylinder to have the same surface area as the droplet, which is the relevant parameter in the context of interfacial tension and diffusion. Assuming this cylinder to be surrounded by a constant interfacial-tension gradient, we 
\begin{figure}[ht!]
    \centering
    \includegraphics[width=8.6cm]{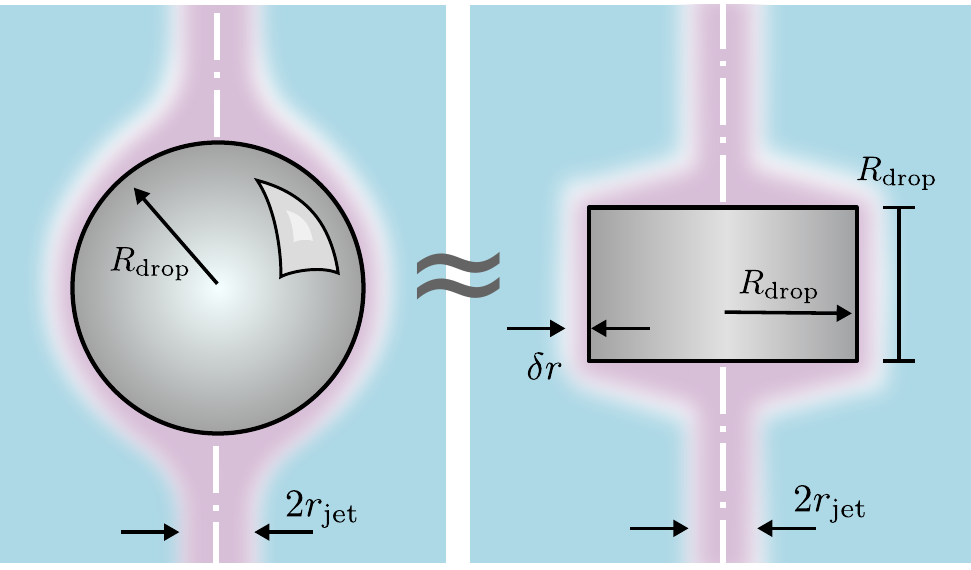}
    \caption{
            Sketch of the model for the radial diffusion around the droplet. On the left the spherical droplet is shown. On the right, the cylindrical surrogate droplet is shown, where the thickness of the ethanol layer around the droplet is denoted as $\delta{}r$.
        }
        \label{fig:kinematics}
\end{figure}
write the Marangoni force acting in negative $z$-direction as
\begin{equation}
    F_{\mathrm{Marangoni}} = 2\pi{}R_{\mathrm{drop}}\Delta\gamma,
    \label{eq:FMarangoni}
\end{equation}
where $\Delta\gamma$ denotes the difference in interfacial tension between the top and the bottom of the cylindrical surrogate droplet. Equating the drag force and the surface force we obtain
\begin{equation}
    1 = \left.\frac{\mathrm{d}\gamma}{\mathrm{d}w}\right|_{w=\bar{w}}\frac{\Delta{}w}{3\mu\beta{}u_{\mathrm{c}}},
    \label{eq:balance}
\end{equation}
where $\Delta{}w$ is the difference in ethanol mass fraction between the top and the bottom, and $\bar{w}$ is the mean ethanol mass fraction at the surface of the drop. Mass conservation for the purely radial diffusion problem around the periphery of the cylindrical droplet gives
\begin{equation}
    \Delta{}w = D\frac{1}{\delta{}r}\frac{\mathrm{d}w}{\mathrm{d}r}\Delta{}t,
    \label{eq:deltaw}
\end{equation}
where $\delta{}r$ is the thickness of the ethanol layer around the cylindrical droplet and $\Delta{}t$ is the duration
that a fluid element needs to travel from the bottom to the top, which is
\begin{equation}
    \Delta{}t = \frac{R_{\mathrm{drop}}}{\beta{}u_{\mathrm{c}}}.
    \label{eq:Deltat}
\end{equation}
Next, we approximate the radial gradient in the ethanol mass fraction as
\begin{equation}
    \frac{\mathrm{d}w}{\mathrm{d}r} = -\frac{w_{\mathrm{EtOH}}(z_{\mathrm{drop}})|_{{r=0}}}{\delta{}r}.
    \label{eq:dwdr}
\end{equation}
The thickness of the ethanol layer around the cylindrical droplet can be expressed as (see supplementary information \S~S3.1)
\begin{equation}
    \delta{}r = \sqrt{\frac{r_{\mathrm{jet}}^{2}}{\beta}+R_{\mathrm{drop}}^{2}} - R_{\mathrm{drop}},
    \label{eq:deltar}
\end{equation}
where the jet radius $r_{\mathrm{jet}}$ is the sum of the initial radius and the diffusive length
\begin{equation}
    r_{\mathrm{jet}} = r_{0} + \sqrt{\frac{2Dz}{u_{\mathrm{c}}}}.
    \label{eq:rjet}
\end{equation}
While the thickness of the ethanol layer does in general vary along the droplet surface, here we assume $\delta r$ to be constant, which can be interpreted as a zeroth-order approximation with respect to the coordinate along the droplet surface. Inserting equations (\ref{eq:wcenter2}) and (\ref{eq:deltaw}--\ref{eq:rjet}) into equation \eqref{eq:balance} and using the same material models as in the simulations as well as $\beta=1$, we can numerically solve for $R_{\mathrm{drop}}$. The result is plotted as the solid black line in Fig.~\ref{fig:comparison} for $z_{\mathrm{drop}} = \qty{6.5}{\milli\metre}$. For varying axial positions $z_\mathrm{drop}$ the result is plotted in Fig.~\ref{fig:z-over-Q-contour}. In both cases, the centerline velocity is approximated as $u_{\mathrm{c}}=\qty{1.46e-3}{\metre\per\second}$ based on the assumption of a purely pressure-driven flow for ${Q_{\mathrm{total}} = \qty{175}{\micro\litre\per\minute}}$. Again, the actual centerline velocity can be larger than this approximated value due to the buoyancy of the ethanol jet (see supplementary information \S~S3.3). However, $u_{\mathrm{c}}$ does not affect the two limiting cases that correspond to the two branches that we can see in Fig.~\ref{fig:comparison}. In both of these two cases we assume that the droplet radius is large compared to the jet radius
\begin{equation}
    R_{\mathrm{drop}} \gg  r_{\mathrm{jet}}.
\end{equation}
The solution shown in Fig.~\ref{fig:z-over-Q-contour} does not depend, qualitatively and order-of-magnitude-wise, on the (arbitrary) choice of $\beta$ as long as $\beta\sim 1$. 
\begin{figure}[h]
    \centering
    \vspace{0.5cm}
    \includegraphics[width=8.6cm]{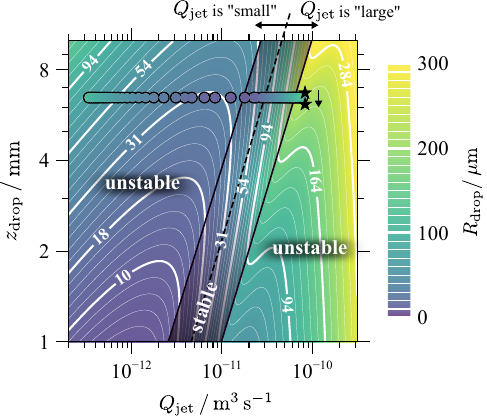}
    \caption{
            Equilibrium configurations of a hovering droplet. The vertical axis shows the droplet position $z_{\mathrm{drop}}$, the horizontal axis shows the jet flow rate $Q_{\mathrm{jet}}$, and the colors/contours indicate the radius $R_{\mathrm{drop}}$ of a droplet that is in equilibrium according to equations (\ref{eq:deltaw}--\ref{eq:rjet}) and equation \eqref{eq:balance}. The equilibrium configurations that are stable according to equation \eqref{eq:stable_unstable} are enclosed by the two solid black lines. Outside of this area, the equilibrium is unstable. The black dashed line indicates where the non-dimensional jet flow rate is unity (${Q_{\mathrm{jet}}/(4\pi{}D_{\mathrm{min}}z_{\mathrm{drop}})=1}$). The circular markers show the results from the numerical simulations (${z_{\mathrm{drop}}=\qty{6.5}{\milli\metre}}$), where the color of the marker represents the radius $R_{\mathrm{drop}}$ of the stationary droplet. The stars correspond the upstream motion (arrow) in the experiment shown in Fig. 3 (between the 4th frame and last frame) for general orientation only, without the intent to imply a quantitative comparison.
        }
        \label{fig:z-over-Q-contour}
        \vspace{0.5cm}
\end{figure}
Next, we calculate the limiting cases for small flow rates (see supplementary information \S~S3.2)
\begingroup
\addtolength{\jot}{0.5em}
\begin{align}
    &Q_{\mathrm{jet}} \ll 4\pi{}D_{\mathrm{max}}z_{\mathrm{drop}} \nonumber \\
    \implies &Q_{\mathrm{jet}}^{\mathrm{crit}} = \left(\left.-\frac{\mathrm{d}\gamma}{\mathrm{d}w}\right|_{w=\bar{w}}\right)^{-1} \frac{12\,\mu\,D_{\mathrm{max}}^{2}z_{\mathrm{drop}}^{3}\pi}{R_{\mathrm{drop}}^{3}}
    \label{eq:smallQ}
\end{align}
and for large flow rates
\begin{align}
    &Q_{\mathrm{jet}} \gg 4\pi{}D_{\mathrm{min}}z_{\mathrm{drop}} \nonumber \\
    \implies &Q_{\mathrm{jet}}^{\mathrm{crit}} = \left(\left.-\frac{\mathrm{d}\gamma}{\mathrm{d}w}\right|_{w=\bar{w}}\frac{4\pi^{2}{}D_{\mathrm{min}}R_{\mathrm{drop}}^{3}}{3\mu}\right)^{1/2}.
    \label{eq:largeQ}
\end{align}
\endgroup
These two limiting cases are indicated in Fig.~\ref{fig:comparison} as thin dashed lines. Note that equations (\ref{eq:smallQ}) and (\ref{eq:largeQ}) are independent of $u_{\mathrm{c}}$ and $\beta$. In retrospect, this justifies the assumption that $u_{\mathrm{c}}$ is approximated by assuming a purely pressure-driven flow that only affects the proposed equilibrium configuration for intermediate jet flow rates. For a more detailed outline of the semi-analytical model we refer the reader to the supp. information \S~S3. \\
\subsection{Axial stability of the hovering droplet}
From the numerical simulations we know that within the hatched area in Fig.~\ref{fig:comparison} we obtain upstream motion of the droplet, and downstream motion otherwise. However, this information does not allow conclusions on the stability of the equilibrium under small variations of the axial position, since the numerical simulations cover only a single value $z_{\mathrm{drop}}$. Instead, we can use the dependence of the semi-analytical model on $z_{\mathrm{drop}}$ to determine the stability: the equilibrium will be stable, if and only if a deflected droplet is pushed back to its original position
\begin{equation}
\left.\frac{\mathrm{d}}{\mathrm{d}z_{\mathrm{drop}}}\left(\frac{F_{\mathrm{Marangoni}}}{F_{\mathrm{drag}}}\right)\right|_{z_{\mathrm{drop}}=z_{\mathrm{drop}}^{\mathrm{crit}}}
\left\{
  \begin{alignedat}{2}
    & > 0 \hspace{0.1cm} \iff \hspace{0.1cm} \mathrm{stable} \\
    & < 0 \hspace{0.1cm} \iff \hspace{0.1cm} \mathrm{unstable,}\label{eq:stable_unstable}
  \end{alignedat}
\right.
\end{equation}
where $z_{\mathrm{drop}}^{\mathrm{crit}} = z_{\mathrm{drop}}$ for $F_{\mathrm{Marangoni}} = F_{\mathrm{drag}}$. In the stable equilibrium case, droplets that are deflected slightly in downstream direction are pushed back to the equilibrium position due to the increased Marangoni force, and droplets that are deflected in upstream direction are pushed back due to the increased drag force. Using the central difference method, we evaluate equation~(\ref{eq:stable_unstable}) numerically. The result is shown in Fig.~\ref{fig:z-over-Q-contour}, revealing that only some of the equilibrium configurations for intermediate flow rates are stable, while configurations for very small and very large flow rates are unstable. This means that hovering droplets can practically only be observed within a limited interval of flow rates. In this regime, the solution shown in Fig.~\ref{fig:z-over-Q-contour} implies that in a quasi-static scenario where we keep the jet flow rate constant, droplet growth will result in an upstream motion (negative $z$-direction). This compares well to what we observe in the experiments. Note that the solid black lines in Fig.~\ref{fig:z-over-Q-contour}, which bound the stable equilibrium configurations, intersect the contour lines (${R_{\mathrm{drop}}=\mathrm{const.}}$) exactly where their tangents are vertical (parallel to the $z$-direction). At these points the equilibrium is independent of $z_{\mathrm{drop}}$, meaning that these are the points of neutral equilibrium that separate the stable configurations from the unstable configurations. \\\indent
Due to the material properties that (through $w_{\mathrm{EtOH}}$ and equation (\ref{eq:wcenter2})) implicitly depend on $z_{\mathrm{drop}}$, the generality of this equilibrium and stability map is limited. The non-monotonic dependence of $z_{\mathrm{drop}}$ on $Q_{\mathrm{jet}}$ that we see in Fig.~\ref{fig:z-over-Q-contour} in the unstable region for large $Q_{\mathrm{jet}}$ is due to the specific material model for interfacial tension: In the limiting case where $Q_{\mathrm{jet}}^{\mathrm{crit}} \gg 4\pi{}Dz_{\mathrm{drop}}$, the critical flow rate does not depend explicitly on $z_{\mathrm{drop}}$ anymore (\textit{cf.} equation~(\ref{eq:largeQ})). If the equilibrium configuration was actually independent of $z_{\mathrm{drop}}$, we would expect that all of the contour lines (${R_{\mathrm{drop}}=\mathrm{const.}}$) are vertical for large jet flow rates. However, this is not the case, as we can see in Fig.~\ref{fig:z-over-Q-contour}. Therefore, we attribute all deviations from vertical contour lines in Fig.~\ref{fig:z-over-Q-contour} for large jet flow rates to the material model for interfacial tension that implicitly depends on $z_{\mathrm{drop}}$.\\\indent
Based on the axisymmetric model employed in the semi-analytical approach, no direct conclusions on the stability of radial or azimuthal modes are possible. The meandering instability that we observed in the experiments could indicate that in many cases, the axisymmetric configuration is unstable. 
The azimuthal stability of the axisymmetric base solutions can be determined numerically by applying the method outlined by \citet{diddens2024bifurcation}, in which the eigenvalue problem of the system perturbed with an azimuthal mode $m$ is solved.
In our calculations, only when artificially manipulating the interfacial tension in between the phases, we obtained the onset of a meandering instability.
Therefore, at this stage, drawing conclusions on the origin of the meandering instability does not appear to be justified. 
\section{Conclusions and Outlook}\label{sec:conclusion}
A droplet that is suspended in a coaxial flow with radial concentration gradients can defy advection and buoyancy and hover at a fixed axial position due to Marangoni stresses. We demonstrated this in an experiment, in which we used a coaxial flow of ethanol and oil on the inside, and water on the outside. Due to the partial immiscibility of these components, droplets spontaneously form downstream of the jet. Using numerical simulations, we showed that a growing nucleation zone of oil droplets forms around the jet, and that droplets suspended inside the jet can hover at a fixed position for a wide range of flow rates. They can even experience a net force in opposite direction of buoyancy and advection that drives them upstream. In some cases, we obtained an azimuthally unstable configuration, hinting at the fact that the exact material model for interfacial tension might play a pivotal role in this context. We then rationalized the resulting flow-regime map using a semi-analytical model that agrees qualitatively with the numerical simulations. For large and small flow rates, the semi-analytical model simplifies to two explicit expressions demonstrating how the critical jet flow rate depends on the droplet radius. For intermediate flow rates, the droplet is stable with respect to variations of its axial position, implying that this mechanism can be used to capture or direct droplets inside a channel flow. The capturing of these droplets can be selective for specific droplet sizes, depending on the applied jet flow rate that is easily controlled in a practical application.\\\indent
One additional degree of freedom that has not been explored in this context is the variation of the cross-sectional area of outer capillary: forcing the co-axial flow through converging or diverging sections of the outer capillary, the flow velocity and the jet radius can be modulated locally. In principle, a repeated arrangement of such necks could be used to build a microfluidic system to fractionate droplets of different sizes. In the application context of dispersive liquid-liquid microextraction, this could enable a continuous process that is easier to scale up since it removes the need for using individual spin tubes and centrifuging of an emulsion to obtain a macroscopic liquid volume containing the analyte. Instead, a macroscopic oil droplet that has grown to a specific size could simply be retained at a specific downstream position and further processed subsequently. In a more general context, the upstream motion allows for a novel approach on pipetting small amounts of liquids or gases from emulsions or dispersions. A corresponding example is a pipette from which a coaxial flow of water (outer flow) and ethanol (jet flow) discharges into a bath of an aqueous emulsion/dispersion. While droplets (or bubbles) are pulled into the jet flow coming from the pipette, the continuous outer phase is flushed away by the outer flow, leaving a clean sample of the dispersed phase inside the pipette. For dispensing the sample, the jet flow is then switched off. Furthermore, the axial motion inside the jet can have implications in the context of process engineering: droplets and bubbles in a multiphase reactor, e.g., in a bubble column, are often not in chemical equilibrium with the continuous phase. A droplet (or bubble) moving relative to the continuous phase can leave a tail of dissolved species behind \cite{falcone20183d} that interacts with other close-by droplets (or bubbles) and possibly entrains them. This could lead to Marangoni-stress-mediated interactions analogous to the interactions studied in the present work with possible implications for rise velocities, mass transfer and coalescence behavior.\\\indent
As indicated, the stability of the equilibrium position of the droplets is sensitive to the material model for the interfacial tension. While this limits the choice of specific liquids to use for droplet capturing, one could also exploit this fact for developing an experimental method for characterizing interfacial tension. One main challenge for using this in systems that are not of the "Ouzo type" lies in introducing small droplets into the co-flow, without disturbing it too much. It is also worth noting that conceptually the hovering and upstream motion  phenomenon does not require a liquid-liquid system. The same phenomenon can be expected for the case of gas bubbles inside a low interfacial tension liquid jet, surrounded by a high interfacial tension outer liquid. Considering specifically the co-axial Ouzo flow system, other aspects remain to be explored in more detail. These include the detailed process of the formation of the large droplets that detach from the jet, involving complex configurations that sometimes lead to multiple levels of nested droplets, and an explanation for the cause of the azimuthal meandering instability of the oil droplet.

\section*{Author contributions}
S. B.: investigation, methodology, writing - original draft, visualization, data curation D. R.: investigation, software, writing - original draft, visualization, data curation S. D.: investigation, writing - original draft C. D.: investigation, software, writing - original draft T. B.: methodology, writing - original draft D. L.:  conceptualization, writing - original draft S. H.:  conceptualization, project administration, supervision, writing - original draft

\section*{Conflicts of interest}
There are no conflicts to declare.

\section*{Data availability}
Data used in this study will be made available upon request to
the corresponding author.

\section*{Acknowledgements}
The authors thank Joachim Groß, Clemens Hansemann, Alexander May and Leon Schuhmann for helping to establish the experimental setup. Funding by the Deutsche Forschungsgemeinschaft (German Research Foundation), Project ID 455566770 is greatly acknowledged.
This work was also supported by an Industrial Partnership Programme of the Netherlands Organisation for Scientific Research (NWO) \& High Tech Systems and Materials (HTSM), 
co-financed by Canon Production Printing Netherlands B.V., University of Twente, and Eindhoven University of Technology (project TKI HTSM - CANON - P1 - PRINTHEAD \& DROPLET FORMATION, grant no. PPS2107). 

\bibliography{rsc}

\end{document}


\title{\vspace{-1cm}\textbf{Upstream motion of oil droplets in co-axial Ouzo flow due to Marangoni forces\\} \vspace{0.5cm}\textbf{Supplementary Information}}

\author{Steffen Bisswanger\textit{$^{a}$}, Duarte Rocha\textit{$^{b}$}, Sebastian Dehe\textit{$^{a,c}$}, Christian Diddens\textit{$^{b}$},\\Tobias Baier\textit{$^{a}$}, Detlef Lohse\textit{$^{b, d}$}, and Steffen Hardt\textit{$^{a, e}$} }

\date{
\footnotesize{
\begin{itemize}
\item[\textit{$^a$}] \footnotesize{Technische Universität Darmstadt, Fachgebiet Nano- und Mikrofluidik, Peter-Grünberg-Str. 10, 64287 Darmstadt, Germany}
\item[\textit{$^b$}] \footnotesize{Physics of Fluids Group and Max Planck Center for Complex Fluid Dynamics, Department of Science and Technology, J.M. Burgers Center for Fluid Dynamics, University of Twente, Enschede, The Netherlands}
\item[\textit{$^c$}] \footnotesize{Linac Coherent Light Source SLAC National Accelerator Laboratory, Menlo Park, California 94025, United States}
\item[\textit{$^d$}] \footnotesize{Max Planck Institute for Dynamics and Self-Organization, Am Faßberg 17, Göttingen, Germany}
\item[\textit{$^e$}] \footnotesize{hardt@nmf.tu-darmstadt.de}
\end{itemize}
}
}
\maketitle
\section{Radial migration of nanodroplets}
\begin{figure}[h!]
    \centering
    \includegraphics[width=14.1cm]{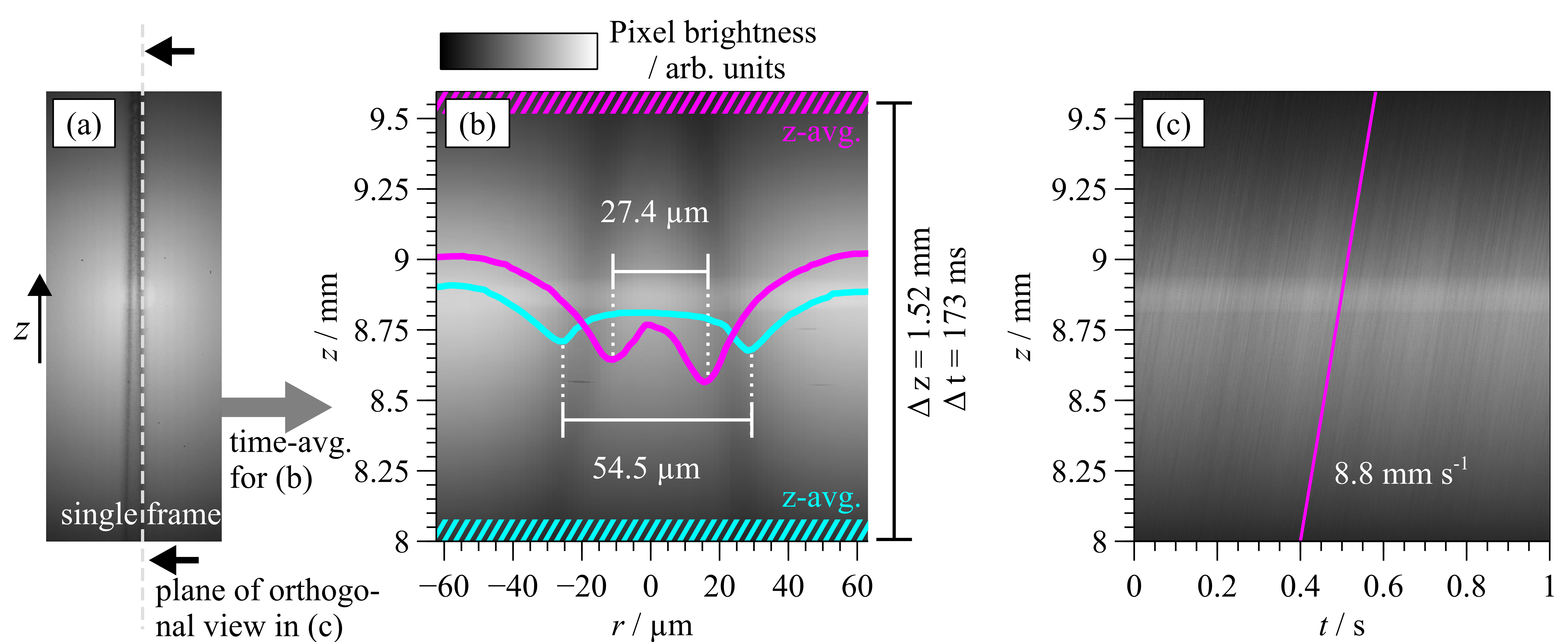}
    \caption{\label{fig:rad_motion} (a): Single frame of the Ouzo jet. (b): Time-averaged pixel brightness value of the region around the Ouzo jet as a function of the vertical coordinate $z$ and the radial coordinate $r$ (not scaled equally). The cyan and magenta lines correspond to a spatial average of the radial brightness distribution (in arbitrary brightness units) at the bottom and top of the frame, as indicated by the hatched areas colored in cyan and magenta. The height of the frame is $\Delta z$=\qty{1.52}{\milli\metre}, which corresponds to a liquid residence time inside the frame of $\Delta t=\qty{173}{\milli\second}$ at a jet velocity of $\qty{8.8}{\milli\metre\per\second}$. (c): Orthogonal view of the side of the jet giving the axial position $z$ as a function of time $t$. The brightness streaks correspond to the movement of individual droplets. The magenta line that is parallel to the streaks indicates a velocity in $z$-direction of $\qty{8.8}{\milli\metre\per\second}$.}
\end{figure}

By means of image analysis (see Fig.~\ref{fig:rad_motion}) we can extract an estimation for the average speed of the radially migrating nanodroplets for the stationary Ouzo jet, that is
\begin{equation}
    \frac{\qty{54.5}{\micro\metre}-\qty{27.4}{\micro\metre}}{2\cdot\qty{173}{\milli\second}} = \qty{78}{\micro\metre\per\second}.
\end{equation}
If the jet was accelerating in axial direction this would explain the radial inward motion of the droplets due to mass conservation. However, since the axial jet velocity is constant, as shown in Fig.~\ref{fig:rad_motion} (c), this radial motion is relative to the surrounding continuous phase. 
\newpage
\section{Flow around droplet in the two-phase model}

\begin{figure}[h!]
    \centering
    \includegraphics[width=0.8\textwidth]{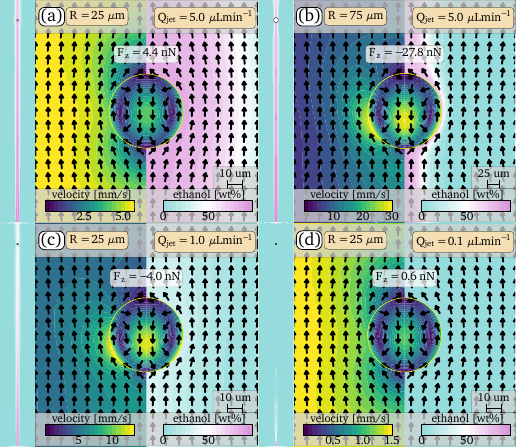}
    \caption{Ethanol concentration field (left part of each subfigure) over the whole domain as well as flow and concentration field around the droplet (right part of each subfigure) for four representative combinations of droplet radius and ethanol jet flow rate, as defined by the two-phase model in \S~III~C: (a) $R_{\mathrm{drop}}$ = \SI{25}{\micro\metre} and $Q_\text{jet}$ = \SI{5}{\micro\liter\per\minute}, (b) $R_{\mathrm{drop}}$ = \SI{75}{\micro\metre} and $Q_\text{jet}$ = \SI{5}{\micro\liter\per\minute}, (c) $R_{\mathrm{drop}}$ = \SI{25}{\micro\metre} and $Q_\text{jet}$ = \SI{1}{\micro\liter\per\minute}, (d) $R_\mathrm{drop}$ = \SI{25}{\micro\metre} and $Q_\text{jet}$ = \SI{0.1}{\micro\liter\per\minute}. In each subfigure, the net $z$-force (positive downstream) is shown.
    }
    \label{fig:droplet_size}
\end{figure}

Figure \ref{fig:droplet_size} illustrates both the ethanol concentration field within the entire geometry (left side of each subfigure) as well as the flow and concentration field surrounding the droplet (right side of each subfigure) for four representative combinations of droplet radius and ethanol jet flow rate. These parameter sets span the phase space defined by the two-phase model discussed in \S~III~C.

\noindent{}At the experimental ethanol flow rate of \SI{5}{\micro\liter\per\minute}, a droplet with a radius of \SI{25}{\micro\metre} is dominated by advective forces and moves upward, Fig. \ref{fig:droplet_size} (a). This occurs because the ethanol jet fully encapsulates the droplet surface, largely eliminating compositional gradients and thereby suppressing Marangoni stresses.
In contrast, a larger droplet with a radius of \SI{75}{\micro\metre} extends beyond the ethanol jet width, allowing for significant compositional gradients to develop at the interface, Fig. \ref{fig:droplet_size} (b). 
These gradients generate Marangoni stresses that drive the flow in the opposite direction, resulting in an upstream (downward) motion of the droplet.
If the droplet radius is kept at \SI{25}{\micro\metre} but the ethanol flow rate is reduced, the encapsulation weakens, allowing compositional gradients to form, Fig. \ref{fig:droplet_size} (c). This reintroduces Marangoni effects, producing a net downward force. However, further reduction of the flow rate causes the droplet surface to be surrounded primarily by water, again eliminating the gradients necessary for Marangoni-driven flow, Fig. \ref{fig:droplet_size} (d).

\section{Additional notes on deriving the semi-analytical model}
\subsection{Ethanol layer around the droplet}
As the ethanol flows around the droplet, the flow is squished in radial direction and dilated in flow direction. The resulting thickness $\delta{}r$ of the ethanol layer around the cylindrical surrogate droplet can be approximated by a mass balance
\begin{equation}
    \int_{\mathrm{jet}} u(r,z) \,\mathrm{d}A = \pi{}r_{\mathrm{jet}}^2{}u_{\mathrm{c}} = \pi\left( \left( R_{\mathrm{drop}}+\delta{r} \right)^{2}- R_{\mathrm{drop}}^{2} \right)\beta{}u_{\mathrm{c}},
    \label{eq:full_model}
\end{equation}
where we assume that radial velocity gradients are small. Rearranging for the thickness of the ethanol layer we obtain
\begin{equation}
    \delta{}r = \sqrt{\frac{r_{\mathrm{jet}}^{2}}{\beta}+R_{\mathrm{drop}}^{2}} - R_{\mathrm{drop}}.
    \label{eq:deltar}
\end{equation}

\subsection{Full expression of the force balance and limiting cases}
\begin{figure}[ht!]
    \centering
    \includegraphics[width=15cm]{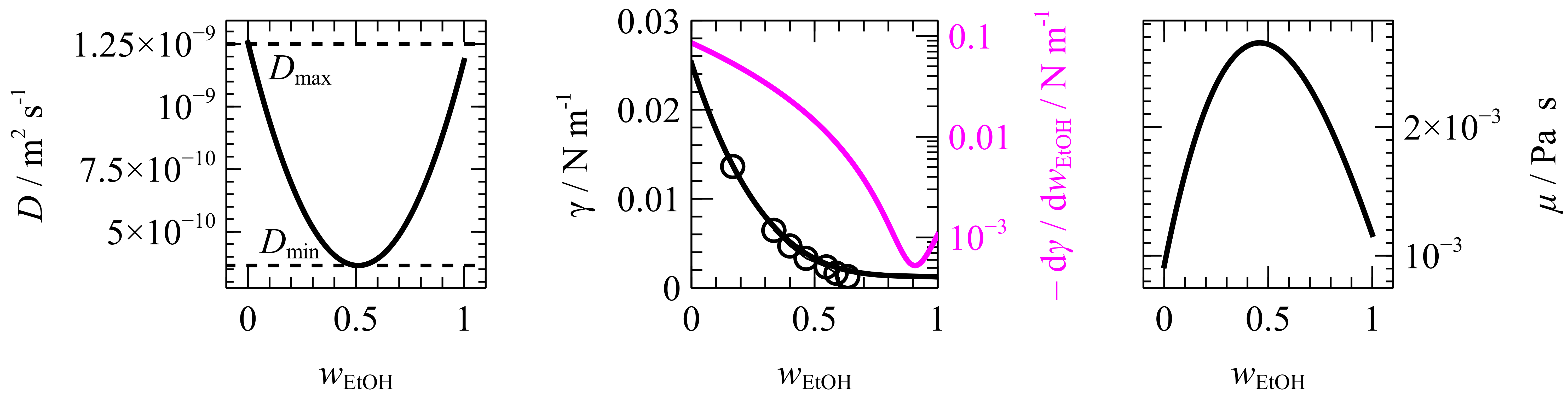}
    \caption{Material models for the diffusion coefficient $D$ taken from Pa\u{r}ez et al.\cite{par2013mutual} (left), for the interfacial tension $\gamma$ and interfacial tension gradient with respect to ethanol mass fraction $\mathrm{d}\gamma/\mathrm{d}w_{\mathrm{EtOH}}$ (middle), and for the viscosity $\mu$, taken from Pa\u{r}ez et al.\cite{par2013mutual} (right), as function of ethanol mass fraction $w_{\mathrm{EtOH}}$. 
    The interfacial tension data are fitted to experimental data (circular markers) by Archer et al.\cite{archer2024experimental} using a monotonic fit.
    }
    \label{fig:mat_model}
\end{figure}

The total expression for the force balance becomes, combining equations (6) and (11-16) from the main text,
\begin{equation}
    1 = \left.-\frac{\mathrm{d}\gamma}{\mathrm{d}w}\right|_{w=\bar{w}} 
    \frac{DR_{\mathrm{drop}}}{3\mu\beta^{2}u_{\mathrm{c}}^{2}} 
    \frac{1 - \exp{\left(\frac{-Q_{\mathrm{jet}}^{\mathrm{crit}}}{4\pi{}Dz_{\mathrm{drop}}}\right)}}{\left(\sqrt{\frac{1}{\beta}\left(\sqrt{\frac{Q_{\mathrm{jet}}^{\mathrm{crit}}}{\pi{}u_{\mathrm{c}}}}+\sqrt{\frac{2Dz_{\mathrm{drop}}}{u_{\mathrm{c}}}}\right)^{2}+R_{\mathrm{drop}}^{2}}-R_{\mathrm{drop}}\right)^{2}}.
    \label{eq:full-model}
\end{equation}
The material models for the interfacial tension gradient $\mathrm{d}\gamma/\mathrm{d}w$, the viscosity $\mu$ and the diffusion coefficient $D$ depend on the mass fraction of ethanol, as shown in Fig.~\ref{fig:mat_model}.
\\\indent
For numerically solving equation~(\ref{eq:full-model}), we approximate these material parameters using the ethanol mass fraction given by
\begin{equation}
    w_{\mathrm{EtOH}}|_{{r=0}} = 1 - \exp{\left(\frac{-Q_{\mathrm{jet}}}{4\pi{}Dz_{\mathrm{drop}}}\right)}.
    \label{eq:wcenter3}
\end{equation}
For the case that $w_{\mathrm{EtOH}} \geq 0.5$, we constrain the diffusion coefficient to $D=D_{\mathrm{min}}$, due to the fact that the radial diffusion problem is dominated by the minimum diffusion coefficient along the radial direction. In order to simplify equation (\ref{eq:full-model}) we now assume that
\begin{equation}
    R_{\mathrm{drop}}^{2} \gg \frac{1}{\beta}\left(\sqrt{\frac{Q_{\mathrm{jet}}}{\pi{}u_{\mathrm{c}}}}+\sqrt{\frac{2Dz_{\mathrm{drop}}}{u_{\mathrm{c}}}}\right)^{2} \equiv \frac{r_{\mathrm{jet}}^{2}}{\beta}.
\end{equation}
The first-order Taylor expansion of equation~\eqref{eq:deltar} around $r_{\mathrm{jet}}^{2} / \beta = 0$ then gives
\begin{equation}
    \delta{}r = \sqrt{\frac{r_{\mathrm{jet}}^{2}}{\beta} + R_{\mathrm{drop}}^{2}} - R_{\mathrm{drop}} = \frac{r_{\mathrm{jet}}^{2}}{2\beta{}R_{\mathrm{drop}}}
\end{equation}
so that equation (\ref{eq:full-model}) becomes
\begin{equation}
    1 = \left.-\frac{\mathrm{d}\gamma}{\mathrm{d}w}\right|_{w=\bar{w}} 
    \frac{4DR_{\mathrm{drop}}^{3}}{3\mu} 
    \frac{1 - \exp{\left(\frac{-Q_{\mathrm{jet}}^{\mathrm{crit}}}{4\pi{}Dz_{\mathrm{drop}}}\right)}}{\left(\sqrt{\frac{Q_{\mathrm{jet}}^{\mathrm{crit}}}{\pi{}}}+\sqrt{2Dz_{\mathrm{drop}}}\right)^{4}}.
    \label{eq:small-drop-jet-ratio-case}
\end{equation}
If we assume that $Q_{\mathrm{jet}} \ll 4\pi{}Dz_{\mathrm{drop}}$, we get
\begin{equation}
    1 = \left.-\frac{\mathrm{d}\gamma}{\mathrm{d}w}\right|_{w=\bar{w}} 
    \frac{4DR_{\mathrm{drop}}^{3}}{3\mu} 
    \frac{\frac{Q_{\mathrm{jet}}^{\mathrm{crit}}}{4\pi{}Dz_{\mathrm{drop}}}}{\left(\sqrt{2Dz_{\mathrm{drop}}}\right)^{4}} 
    \qquad \implies \qquad 
    Q_{\mathrm{jet}}^{\mathrm{crit}} = \left(-\left.\frac{\mathrm{d}\gamma}{\mathrm{d}w}\right|_{w=\bar{w}}\right)^{-1} \frac{12\,\mu\,D^{2}z_{\mathrm{drop}}^{3}\pi}{R_{\mathrm{drop}}^{3}}.
    \label{eq:smallQ}
\end{equation}
On the other hand, if we assume that $Q_{\mathrm{jet}} \gg 4\pi{}Dz_{\mathrm{drop}}$, we get
\begin{equation}
    1 = \left.-\frac{\mathrm{d}\gamma}{\mathrm{d}w}\right|_{w=\bar{w}} 
    \frac{4DR_{\mathrm{drop}}^{3}}{3\mu} 
    \frac{1}{\left(\sqrt{\frac{Q_{\mathrm{jet}}^{\mathrm{crit}}}{\pi{}}}\right)^{4}} 
    \qquad\implies\qquad 
    Q_{\mathrm{jet}}^{\mathrm{crit}} = \left(-\left.\frac{\mathrm{d}\gamma}{\mathrm{d}w}\right|_{w=\bar{w}}\frac{4\pi^{2}{}DR_{\mathrm{drop}}^{3}}{3\mu}\right)^{1/2}.
    \label{eq:largeQ}
\end{equation}
\subsection{Centerline velocity and buoyancy effects}
\begin{figure}[ht!]
    \centering
    \includegraphics[width=15cm]{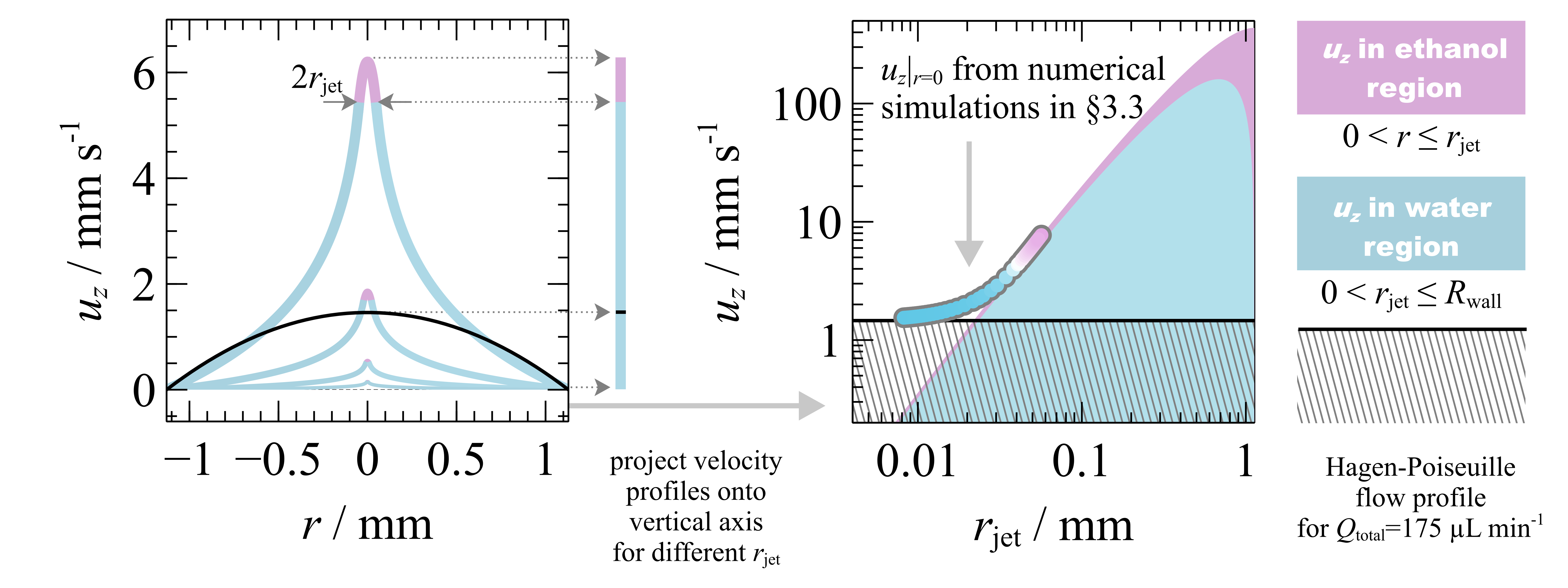}
    \caption{Axial flow velocity of a channel flow driven by a buoyant jet according to equations (\ref{eq:uz_jet}) and (\ref{eq:uz_water}) for $\mu = \qty{1.44}{\milli\pascal\second}$. Left: $u_z$ as a function of the radial coordinate for four different values of $r_{\mathrm{jet}} = \left\{6.25, 12.5, 25, 50\right\}\qty{}{\micro\metre}$. The blue sections of the curves indicate the water region, the pink sections indicate the buoyant ethanol region. The black line shows a Hagen-Poiseuille flow profile for $Q_{\mathrm{total}}=\qty{175}{\micro\litre\per\minute}$ for reference. Right: $u_z$ as function of the jet radius $r_{\mathrm{jet}}$. The plot on the right represents the flow profiles from the plot on the left, projected onto the vertical axis for arbitrary values of $r_{\mathrm{jet}}$. The hatched region indicates the Hagen-Poiseuille flow velocities for reference. The blue and pink area indicate the water and the ethanol region. For comparison, the centerline velocities of the numerical simulations from \S~III~C at $z=z_{\mathrm{drop}}-2R_{\mathrm{drop}}$ that include diffusion effects are shown. For the numerical results $r_{\mathrm{jet}}:=\sqrt{Q_{\mathrm{jet}}/\left(\pi{}u_{\mathrm{c}}\right)}$. The comparison shows that for small $r_{\mathrm{jet}}$ (\textit{i.e.}, small $Q_{\mathrm{jet}}$) the flow profile in the numerical simulation is dominated by the pressure-driven flow, while for large $r_{\mathrm{jet}}$ (\textit{i.e.}, large $Q_{\mathrm{jet}}$) it is dominated by the buoyancy-induced flow. The colormap of the numerical simulation results indicates $w_{\mathrm{EtOH}}$ in the same way as in Fig.~5 of the main text.}
    \label{fig:etoh-buoyancy}
\end{figure}

Without taking buoyancy into account, we would expect the centerline velocity in the capillary to be approximately
\begin{equation}
    u_{\mathrm{c}} = \frac{2Q_{\mathrm{water}}}{4\,\qty{}{\square\milli\metre}} = \qty{1.46}{\milli\metre\per\second},
\end{equation}
which is the expected flow velocity at the center for a Hagen-Poiseuille flow through a pipe with a circular cross-sectional area of $\qty{4}{\square\milli\metre}$ if we assume that $Q_{\mathrm{water}} = \qty{175}{\micro\litre\per\minute}\gg{}Q_{\mathrm{EtOH}}$. From the simulations we obtain values for $u_{\mathrm{c}}$ that are larger. This is due to the buoyancy of the ethanol jet that speeds up the flow at the center. This means that the assumption of a pressure-driven Hagen-Poiseuille flow looses its validity with increasing ethanol flow rate. To assess the consequences of this, we can consider the following case of a flow that is purely driven by buoyancy:
We solve the Navier-Stokes equation for a pipe flow where the pipe is vertical (parallel to gravitational acceleration) and where we have a discontinuous jump in the mass density between the sheath flow and the jet flow. The governing equations and corresponding boundary conditions are
\begin{equation}
    0 = \mu\left(\frac{1}{r}\frac{\partial}{\partial r}\left(r\frac{\partial u_{z}}{\partial r}\right)\right) - \Delta\rho g
\end{equation}
\begin{equation}
    \left. u_{z}^{\mathrm{sheath}}\right\rvert_{r=R_{\mathrm{wall}}} = 0
\end{equation}
\vspace{3pt}
\begin{equation}
    \left.u_{z}^{\mathrm{sheath}}\right\rvert_{r=r_{\mathrm{jet}}} = \left.u_{z}^{\mathrm{jet}}\right\rvert_{r=r_{\mathrm{jet}}}
\end{equation}
\vspace{3pt}
\begin{equation}
    \left.\frac{\partial u_{z}^{\mathrm{sheath}}}{\partial r}\right\rvert_{r=r_{\mathrm{jet}}} = \left.\frac{\partial u_{z}^{\mathrm{jet}}}{\partial r}\right\rvert_{r=r_{\mathrm{jet}}}
\end{equation}
\vspace{3pt}
\begin{empheq}[left={u_{z}=\empheqlbrace}]{alignat=2}
    & u_{z}^{\mathrm{jet}}         && \text{for}\ 0 < r \leq r_{\mathrm{jet}} \\
    & u_{z}^{\mathrm{sheath}}\quad      && \text{for}\ r_{\mathrm{jet}} < r \leq R_{\mathrm{wall}}
\end{empheq}
and
\begin{empheq}[left={\Delta\rho=\empheqlbrace}]{alignat=2}
    & \qty{200}{\kilo\gram\per\cubic\metre}\quad         && \text{for}\ 0 < r \leq r_{\mathrm{jet}} \\
    & 0                                                  && \text{for}\ r_{\mathrm{jet}} < r \leq R_{\mathrm{wall}}.
\end{empheq}
The solution of this problem is
\begin{empheq}[left={u_{z}=\empheqlbrace}]{alignat=2}
    & \frac{\Delta\rho g}{4\mu}\left(r_{\mathrm{jet}}^{2}\left(1+2\ln{\left(\frac{R_{\mathrm{wall}}}{r_{\mathrm{jet}}}\right)}\right)-r^{2}\right)\quad         && \text{for}\ 0 < r \leq r_{\mathrm{jet}}\label{eq:uz_jet} \\
    & \frac{\Delta\rho g}{2\mu}r_{\mathrm{jet}}^{2}\ln{\left(\frac{R_{\mathrm{wall}}}{r}\right)}                          && \text{for}\ r_{\mathrm{jet}} < r \leq R_{\mathrm{wall}}\label{eq:uz_water}
\end{empheq}
so that
\begin{equation}
    u_{\mathrm{c}} \approx \frac{\Delta\rho g}{2\mu}r_{\mathrm{jet}}^{2}\ln{\left(\frac{R_{\mathrm{wall}}}{r_{\mathrm{jet}}}\right)}.
\end{equation}
These results are plotted in Fig.~\ref{fig:etoh-buoyancy}, in comparison to the purely pressure driven flow that satisfies $Q_{\mathrm{total}}=\qty{175}{\micro\litre\per\minute}$ and in comparison to the numerical simulation results.\\\indent
The flow at the center speeds up as the thickness of the jet increases. For small flow velocities, the estimation based on the Hagen-Poiseuille flow is reasonably good. This is due to the different scaling of the area over which viscous stresses act on the jet ($\propto{}r_{\mathrm{jet}}$) and the buoyancy force that acts on the jet ($\propto{}r_{\mathrm{jet}}^{2}$). Therefore, the solution for the pressure-driven flow dominates the flow profile for small jet flow rates. For large jet flow rates, the centerline velocity can be several times larger than the value obtained from Hagen-Poiseuille flow. However, from equations (\ref{eq:small-drop-jet-ratio-case}) and (\ref{eq:full_model}) we know that the critical flow rate only depends on the centerline velocity, if the droplets are small compared to the jet. This means that the velocity increase due to buoyancy only affects how the two limiting cases (given by equations \eqref{eq:smallQ} and \eqref{eq:largeQ}) transition from one to another, where the critical droplet radius is minimal. This is why the assumption of a constant centerline velocity is adequate for preserving the qualitative dependence of $Q_{\mathrm{jet}}^{\mathrm{crit}}$ on $R_{\mathrm{drop}}$. Additionally, we can draw another conclusion from this model, indicated in Fig. \ref{fig:etoh-buoyancy} (right): The vertical extent of the purple area shows the velocity range over the width of the jet. The fact that this extent is narrow means that the assumption that the centerline velocity is characteristic for the entire jet is adequate.
\bibliography{rsc} 
\bibliographystyle{rsc} 